\documentclass[a4paper,11pt]{article}
\usepackage{pos}
\usepackage{amsmath,graphicx,slashed}

\newcommand{\braket}[1]{\left\langle#1\right\rangle}

\newcommand{\Tr}{\operatorname{Tr}}

\newcommand{\bm}{\boldmath}
\newcommand{\Kslash}{\slashed{K}}
\newcommand{\pslash}{\slashed{p}}
\newcommand{\Dslash}{\slashed{D}}
\newcommand{\Dslashfwd}{\overrightarrow{\Dslash}}
\newcommand{\Dslashbak}{\overleftarrow{\Dslash}}
\newcommand{\bbra}{\left<\!\left<}
\newcommand{\kket}{\right>\!\right>}
\newcommand{\order}{\mathcal{O}}

\newcommand{\gm}{\gamma_\mu}

\title{Quark-gluon vertex with 2 flavours of O(a) improved Wilson fermions}

\author[a]{Ay{\c s}e K{\i}z{\i}lers\"u}
\author[b]{Orlando Oliveira}
\author[b]{Paulo Silva}
\author*[c]{Jon-Ivar Skullerud}
\author[d]{Andr\'e Sternbeck}

\affiliation[a]{CSSM, Department of Physics, Faculty of Sciences, School of Physical Sciences, University of Adelaide, 5005,
Adelaide, Australia.} 

\affiliation[b]{CFisUC, Department of Physics, University of Coimbra, 
3004--516 Coimbra, Portugal.} 

\affiliation[c]{
Department of Theoretical Physics, National University of Ireland Maynooth,
Maynooth, County Kildare, Ireland.}

\affiliation[d]{
  Theoretisch-Physikalisches Institut and Universit\"atsrechenzentrum,
  Friedrich-Schiller-Universit\"at
Jena, 07743 Jena, Germany}

\emailAdd{jonivar.skullerud@mu.ie}

\abstract{
  We study the Landau-gauge quark-gluon vertex with 2 flavours of O(a)
  improved Wilson 
fermions, for several lattice spacings and quark masses. In the limit
of vanishing gluon momentum, we find that all nonzero form factors
have a significant infrared strength, and that the leading form factor
$\lambda_1$, multiplying the tree-level vertex structure, is significantly
enhanced in the infrared compared to the quenched case. We find that
all form factors are further enhanced in the infrared as the chiral
and continuum limits are approached.
}

\FullConference{%
 The 38th International Symposium on Lattice Field Theory, LATTICE2021
  26th-30th July, 2021
  Zoom/Gather@Massachusetts Institute of Technology
}

\begin{document}
\maketitle

\section{Introduction}

The quark--gluon vertex is one of the basic ingredients of QCD, along
with the gluon and quark propagators and the three-gluon and
four-gluon vertices.  It
encodes the fundamental interaction between quarks and gluons, and as
such lends itself naturally to defining an effective charge (running
coupling).  It is also a crucial ingredient in the gap equation
(Dyson--Schwinger equation) for the quark propagator.  Indeed, its
role in this equation and its relation to dynamical chiral symmetry
breaking (D$\chi$SB) and quark confinement provides a compelling
reason to study it in more detail.

Since the nonperturbative quark and gluon propagators (at least in
Landau gauge) are by now very well known, it is clear that consistent
results, including a sufficient level of D$\chi$SB and adequate meson
phenomenology, cannot be obtained with a bare vertex or indeed a
vertex constructed from the abelian Ward--Takahashi identity.  In
particular, vertex structures beyond the one found at tree level are
crucial to obtaining the correct phenomenology.

In recent years there has been significant progress in our understanding of
the quark--gluon vertex using functional methods and constraints from
the longitudinal and transverse Slavnov--Taylor identities (see for
example 
\cite{Williams:2015cvx,Binosi:2016wcx,Aguilar:2018epe,Gao:2021wun,Albino:2021rvj}).  There has only been a very small number of
lattice calculations, mostly in the quenched approximation
\cite{Skullerud:2002ge,Skullerud:2003qu,Lin:2005zd,Kizilersu:2006et}.
Here we report on and update a recent calculation with $N_f=2$
dynamical fermions and several lattice spacings 
and quark masses \cite{Kizilersu:2021jen}.

\section{Calculation details}

\subsection{Lattice parameters}

We have used gauge ensembles from Regensburg QCD (RQCD) collaboration
\cite{Bali:2014gha}, with $N_f=2$ nonperturbatively clover-improved
Wilson fermions.  The parameters used are listed in
Table~\ref{tab:params}.  In addition to the RQCD ensembles, we have
produced a quenched ensemble with lattice spacing $a=0.07\,$fm
(matching the L07 and H07 ensembles), but with a larger valence quark
mass $m_\pi\approx1000$MeV.  The configurations have been fixed to
Landau gauge using an overrelaxation algorithm.  More details can be
found in \cite{Kizilersu:2021jen}.
\begin{table}[thb]
\begin{center}
\begin{tabular}{lcc|cccccc}
Name & $\beta$ & $\kappa$ & $a$ [fm] & $V$ & $m_\pi$ [MeV] & 
 $N_{\text{cfg}}$ & $N_{\text{src}}$ \\ \hline
L08   & 5.20 & 0.13596 & 0.081 & $32^3\times64$ & 280  & 900 & 4 \\
H07   & 5.29 & 0.13620 & 0.071 & $32^3\times64$ & 422  & 900 & 4 \\
L07   & 5.29 & 0.13632 & 0.071 & $32^3\times64$ & 295  & 908 & 4 \\
L07-64& 5.29 & 0.13632 & 0.071 & $64^3\times64$ & 290  & 750 & 2 \\
H06   & 5.40 & 0.13647 & 0.060 & $32^3\times64$ & 426  & 900 & 4 \\\hline
Q07   & 6.16 & 0.1340  & 0.071 & $32^3\times64$ & 1000 & 998 & 4
\end{tabular}
\caption{Lattice ensembles and parameters used in this study.}
\label{tab:params}
\end{center}
\end{table}

We have used the same clover action for the valence quarks as for the
sea quarks, with an off-shell $\order(a)$-improved ``rotated'' quark
propagator,
\begin{equation} 
S_R(x,y) 
 = \braket{(1+b_q am)^2(1-c_qa\Dslashfwd(x))M^{-1}(x,y;U)(1+c_q a\Dslashbak(y))} \; ,
\label{eq:rotprop}
\end{equation}
where $M(x,y;U)$ is the fermion matrix evaluated on a single gauge
configuration $U$.
The quark--gluon vertex is then determined after Fourier transforming by
\begin{align}
\Lambda_\mu^{a,\rm lat}(p,q)
 &= S_R(p)^{-1} \bbra S_R(p;U)A^a_\mu(q)\kket S_R(p+q)^{-1}D(q)^{-1}_{\nu\mu}
\, ,
\label{eq:vtx-amputate}
\end{align}
where $\bbra\cdot\kket$ denotes averaging over gauge
fields only, while $S_R(p;U)$ is the Fourier transform of \eqref{eq:rotprop} evaluated on a
single gauge configuration $U$.

\subsection{Extracting form factors}

The one-particle irreducible quark-gluon vertex, $(\Lambda_\mu^a)_{\beta\rho}^{ij} =
t_{ij}^a\,( -ig_0\,\Gamma_\mu)_{\beta\rho} $, can be written in terms
of 12 independent tensor structures in a general kinematics.  Here we
will consider only the soft gluon kinematics, where the gluon momentum is
zero, and hence the vertex depends only on the quark momentum $p$.
In this kinematics, the quark--gluon vertex can be expressed in terms
of three independent tensor structures with associated form factors $\lambda_{1,2,3}$,
\begin{equation}
  \Gamma_\mu(p) = \lambda_1(p^2)\gamma_\mu
+ 4\lambda_2(p^2)\pslash p_\mu
+ 2i\lambda_3(p^2)p_\mu\,. \label{eq:decomp}
\end{equation}
From \eqref{eq:decomp}, we derive the following expressions which may
be used to determine the form factors $\lambda_i$,
\begin{align}
  \lambda_1
  &= \frac{1}{3}\Big(\delta_{\mu\nu}-\frac{p_\mu p_\nu}{p^2}\Big)
  \Tr_4\gamma_\nu\Gamma_\mu\,,  \label{l1-cov}\\
  \lambda_2
  &= \frac{1}{12p^2}\Big(\delta_{\mu\nu}-4\frac{p_\mu p_\nu}{p^2}\Big)
  \Tr_4\gamma_\nu\Gamma_\mu\,,  \label{l2-cov} \\
  \lambda_3 &= \frac{p_\mu}{2ip^2}\Tr_4\Gamma_\mu\,. \label{l3-cov}
\end{align}
In addition to these \emph{covariant} expressions we also employ the
following \emph{non-covariant} expressions for $\lambda_1,\lambda_2$,
\begin{align}
  \lambda_1 &= \Tr_4\gm\Gamma_\mu\Big|_{p_\mu=0}\,, \label{l1-noncov}\\
  \lambda_2
  &= \frac{p_\mu p_\nu}{4p^2}\Tr_4\gamma_\nu\Gamma_\mu\Big|_{\nu\neq\mu}\,.
  \label{l2-noncov}
\end{align}
In \cite{Kizilersu:2021jen} only the non-covariant expressions were
used.  Here we will also present results using the covariant
expressions.  We note that these were previously used in
\cite{Lin:2005zd}.

\subsection{Tree-level correction}

The tree-level rotated vertex in the soft gluon kinematics is given by
\begin{align}
  \Gamma^{(0)}_{\mu}(p,0,p)
  =  \,\frac{1}{ \big(1+ c_q^2 a^2 K^2(p) \big)^4} 
  & \bigg\{  
         {\boldsymbol{ \gamma_\mu}}\, \Big[ \big(1+c_q^2 a^2K^2(p)\big)^2
           C_\mu(p) \Big] \notag \\
         &  \,\, \bm{-4a^2K_\mu\Kslash(p)}\,
         \Big[ 2c_q^2 C_\mu(p)
           - c_q\big(1-c_q^2a^2K^2(p) \big) \Big] \notag \\
         & \,\,  \bm{-2i aK_\mu}\, \Big[ -2c_q^2 a^2K^2(p)
           +\frac{1}{2}\big(1-c_q^2 a^2K^2(p) \big)\notag \\
         &\qquad  -2c_q \big(1-c_q^2a^2K^2(p)\big)\, C_\mu(p) \Big]
 \bigg\} \,,\label{eq:treelevel}
\end{align}
where the lattice momentum variables $K_\mu, C_\mu$ are given by
\begin{align}
  K_\mu(p) &= \frac{1}{a}\sin(p_{\mu}a)\,,
  & C_\mu(p) &= \cos(p_{\mu}a)\,. \label{def:Kmu-Cmu}
\end{align}
From this we note that all three form factors have a nontrivial
momentum dependence already at tree level on the lattice, and we
correct for this by dividing the raw lattice data for $\lambda_1$
(which is 1 at tree level in the continuum) by its tree-level form,
while for $\lambda_2$ and $\lambda_3$ (which are zero at tree level in
the continuum) we subtract off the tree-level expressions from
\eqref{eq:treelevel}.  The tree-level form also suggests replacing the
Fourier transform momentum $p$ with the lattice momentum $K(p)$ in
\eqref{l1-cov}--\eqref{l2-noncov}.  The detailed procedure for
extracting and tree-level correcting the form factors is presented in
\cite{Kizilersu:2021jen}.

\section{Results}

\begin{figure}[tb]
\begin{center}
  \includegraphics[height=5.2cm]{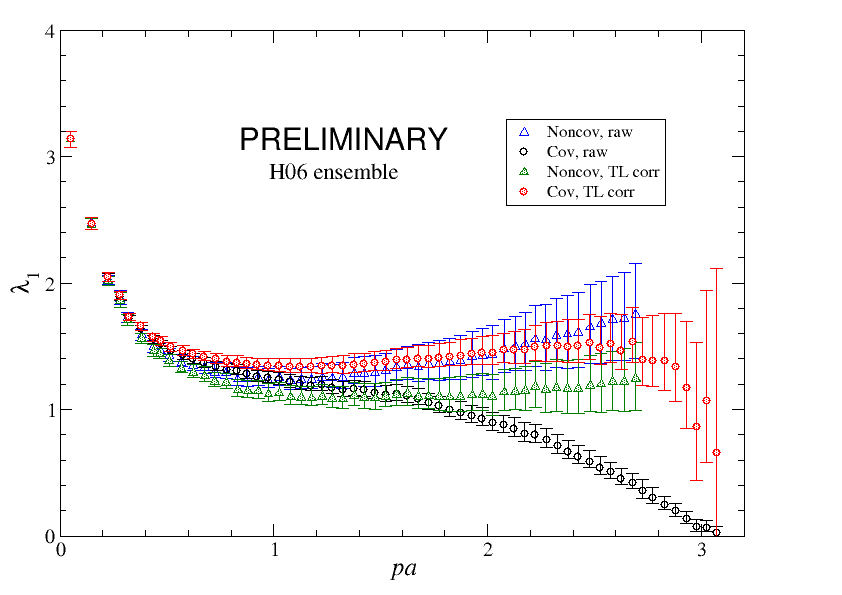}
  \includegraphics[height=5.2cm]{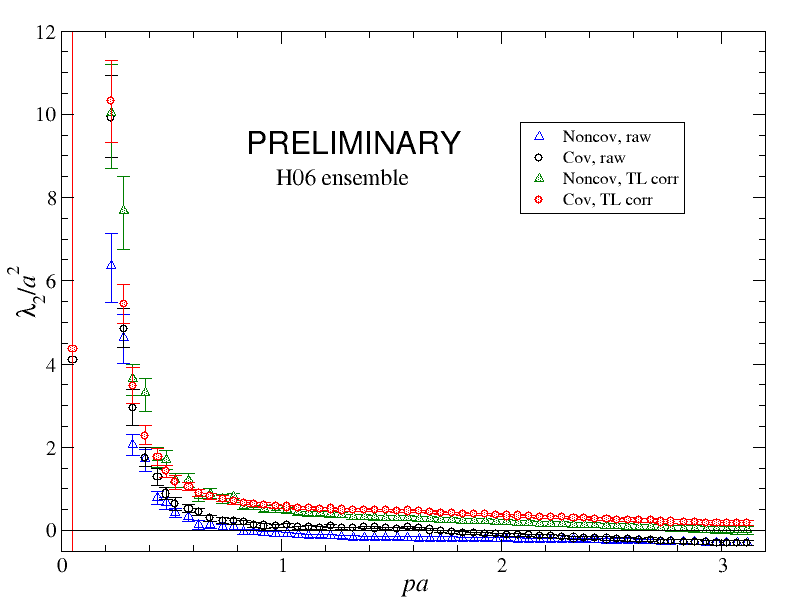}\\
  \includegraphics[height=5.2cm]{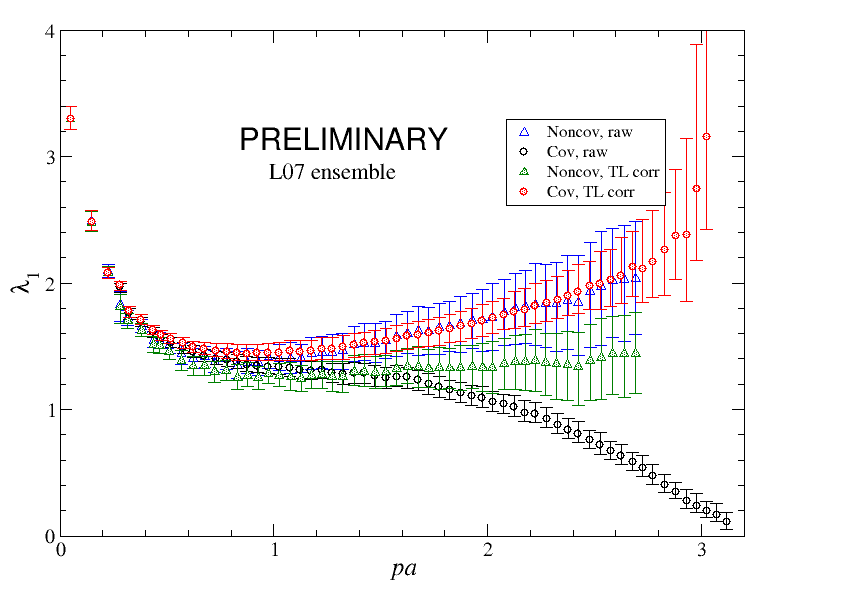}
  \includegraphics[height=5.2cm]{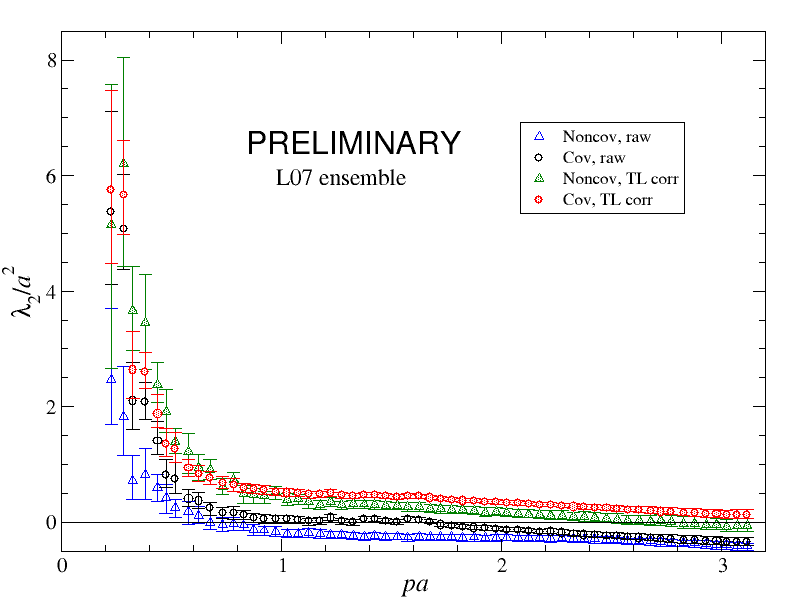}
  \caption{Comparison of covariant and non-covariant determination of
    $\lambda_1$ (left) and $\lambda_2$ (right) from the H06 (top) and
    L07 (bottom) ensembles, with and without tree-level correction.
    The data have not been renormalised.}
  \label{fig:cov-noncov}
\end{center}
\end{figure}
\begin{figure}[tbp]
\includegraphics[width=0.48\textwidth]{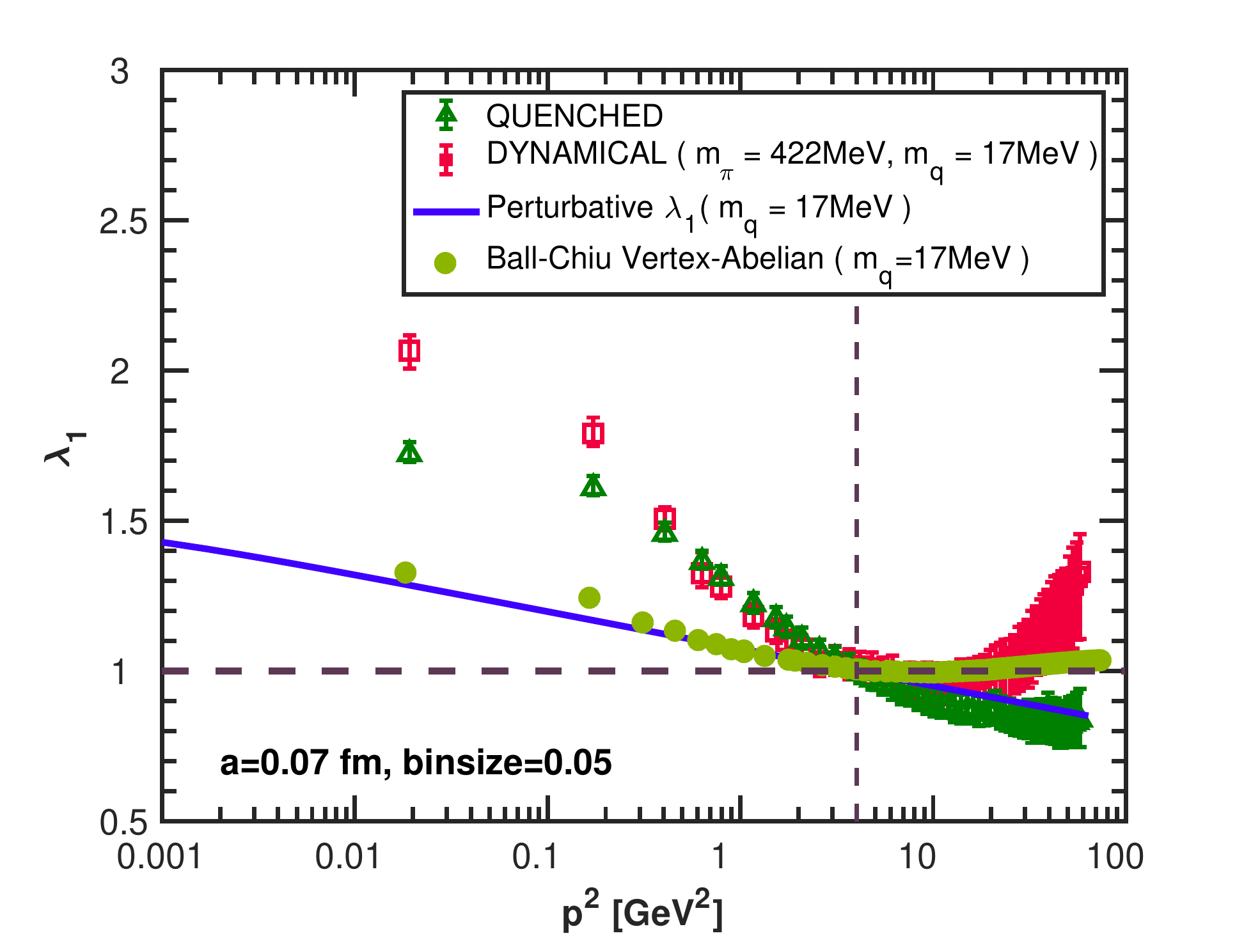}
\includegraphics[width=0.48\textwidth]{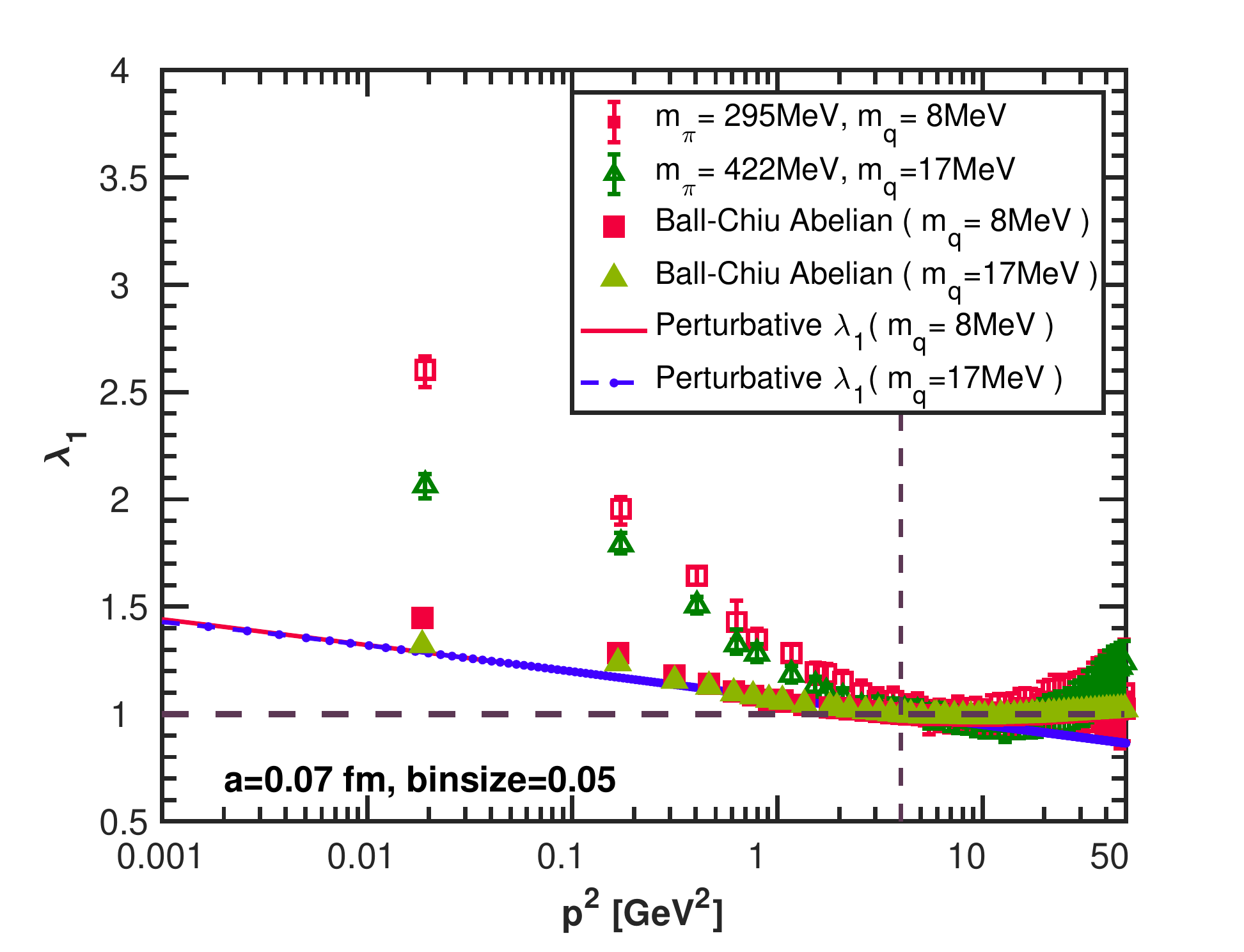}\\
\includegraphics[width=0.48\textwidth]{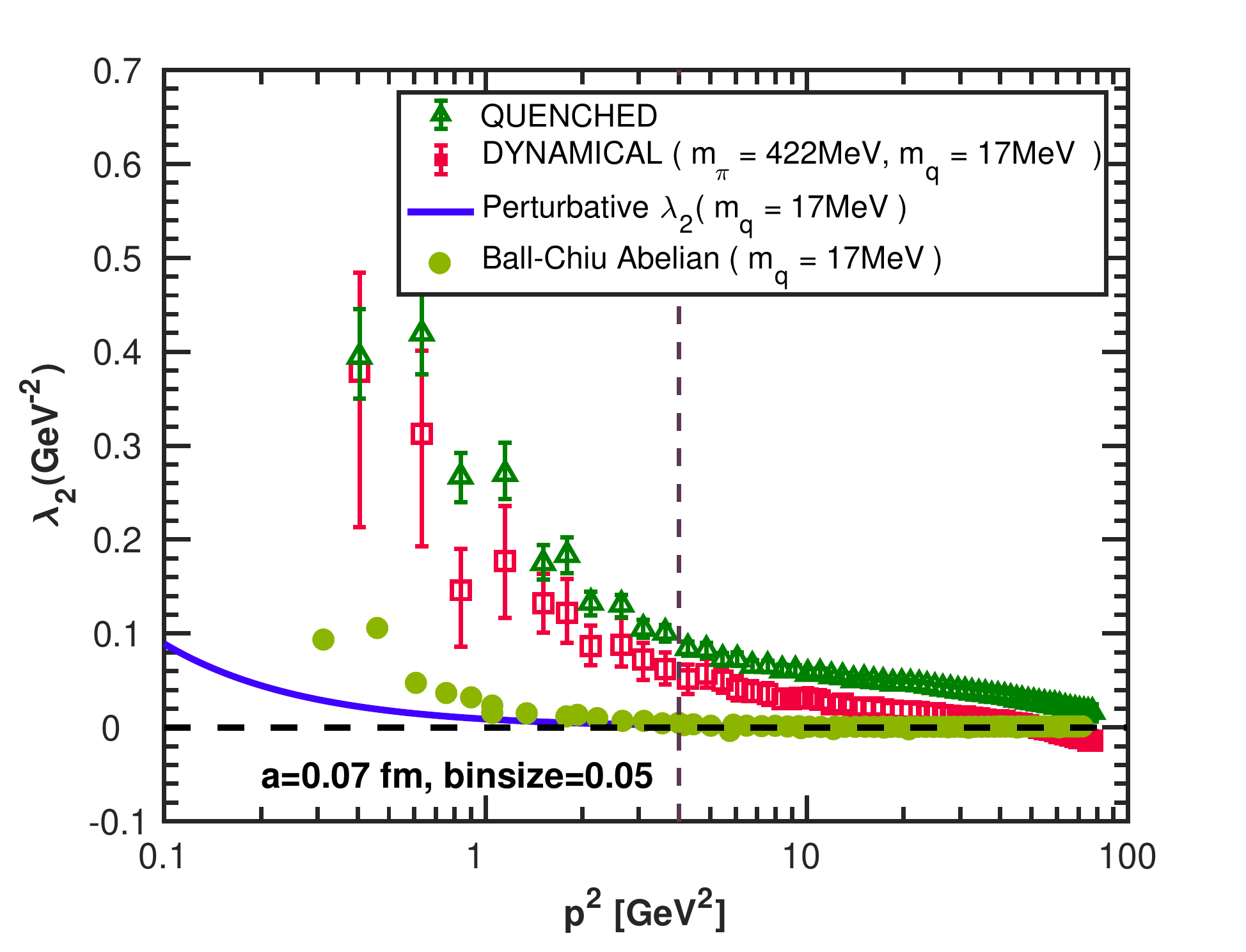}
\includegraphics[width=0.48\textwidth]{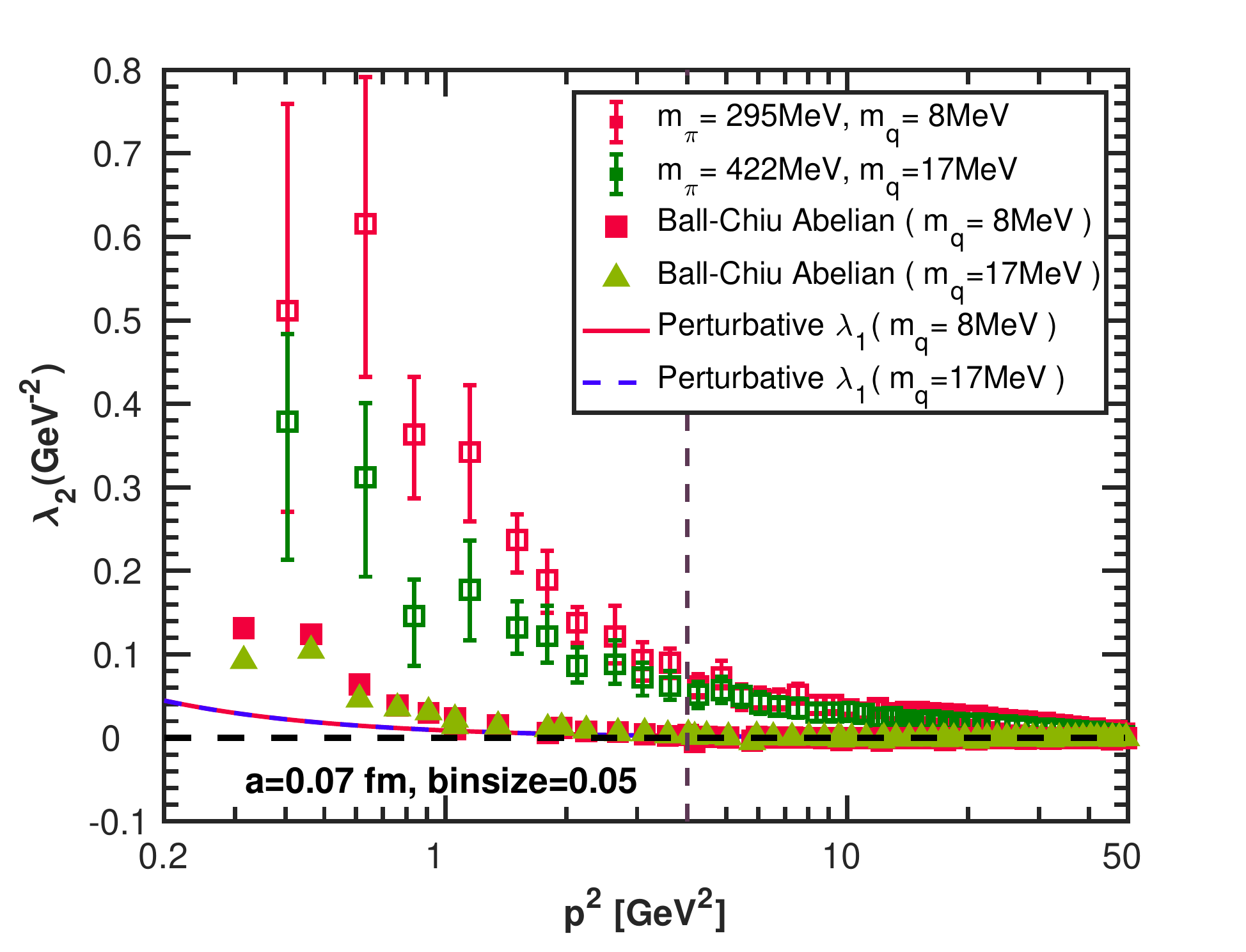}\\
\includegraphics[width=0.48\textwidth]{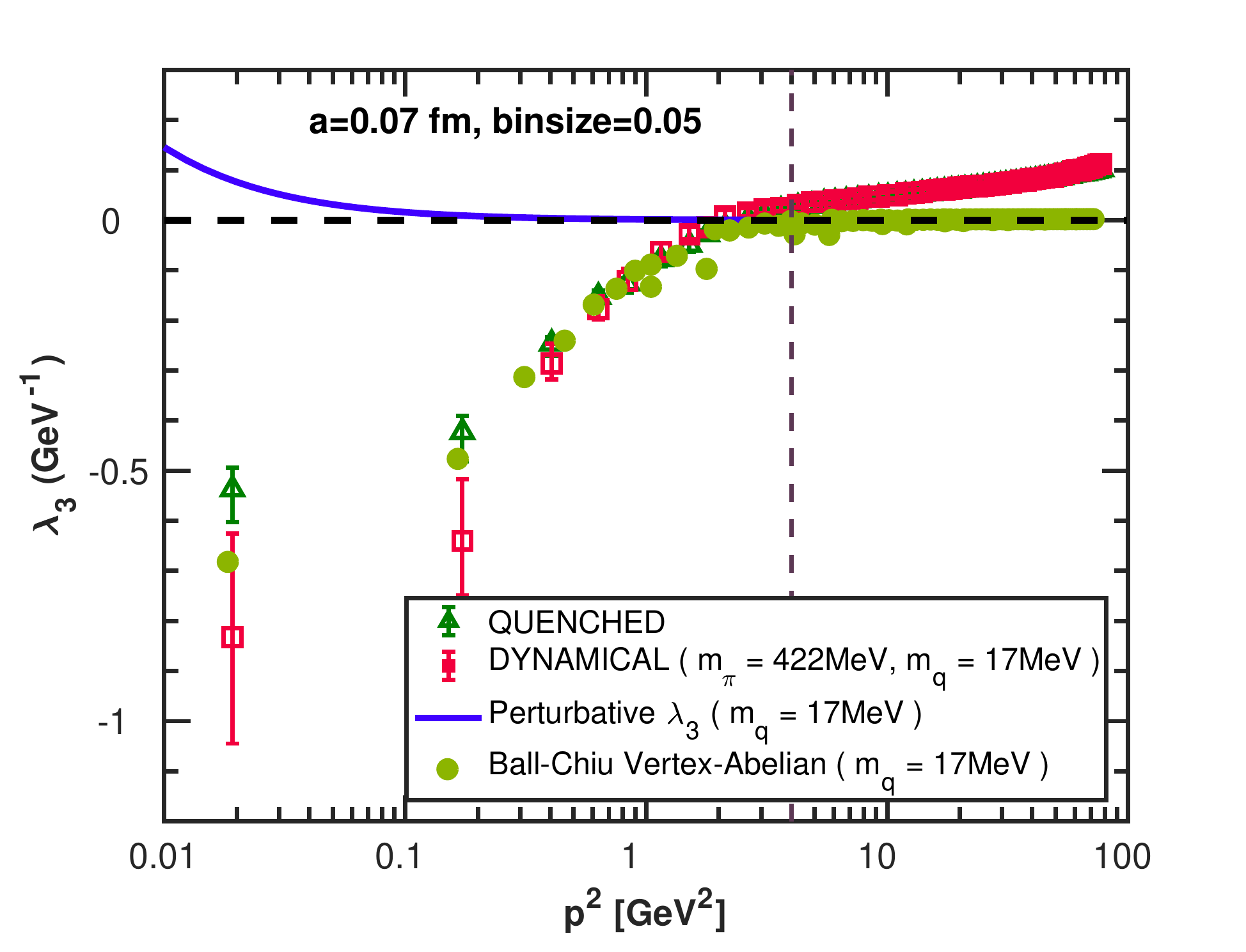}
\includegraphics[width=0.48\textwidth]{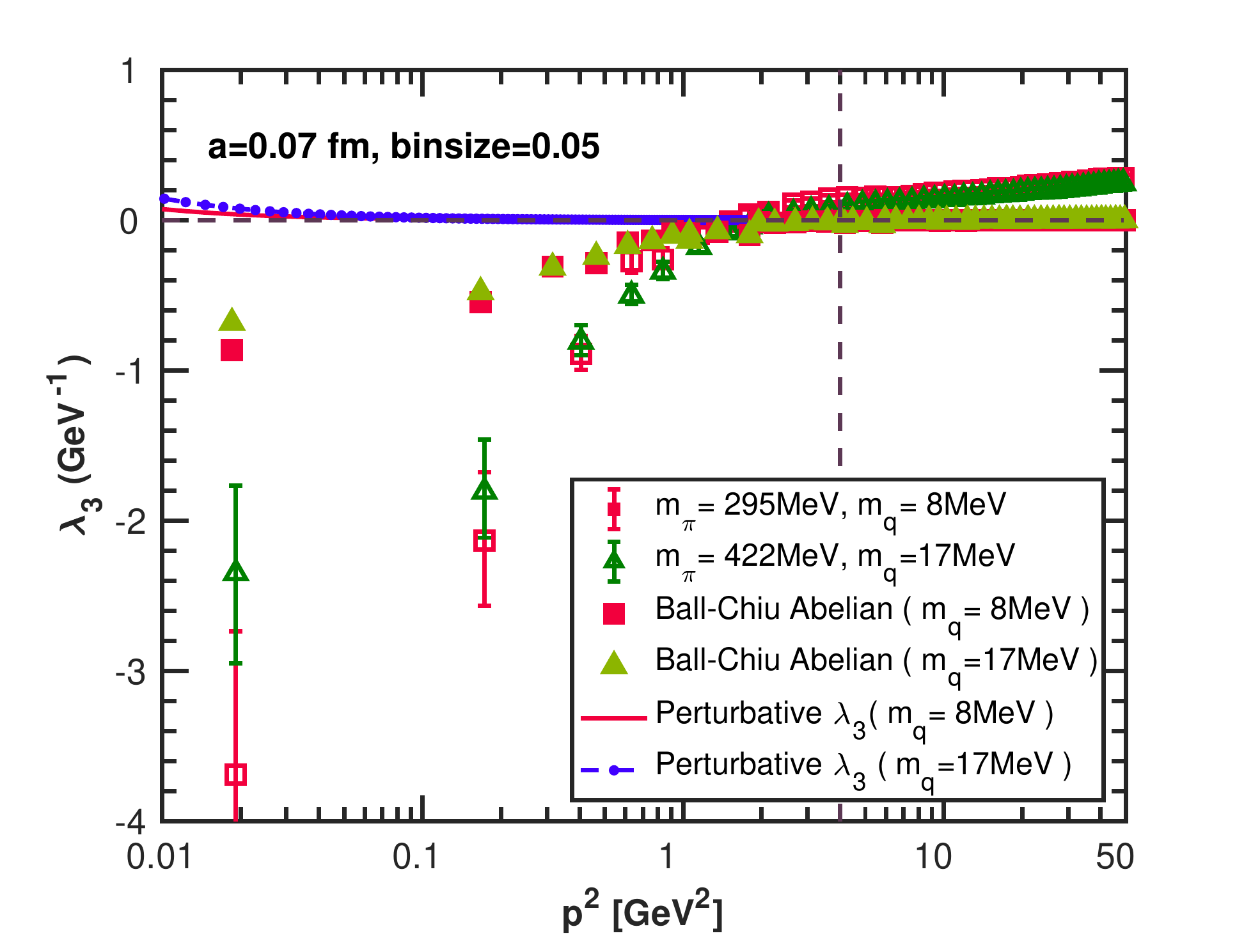}
\caption{Quark mass and flavour dependence of (top to bottom)
  $\lambda_1,\lambda_2$ and $\lambda_3$.  The left panel
  compares the Q07 (quenched) and H07 (dynamical) ensemble, while the
  right panel compares the L07 and H07 ensembles.  Also shown are the
1-loop perturbative result as well as the Ball--Chiu vertex which
satisfies the abelian Ward--Green--Takahashi identity.}
\label{fig:mass-flavour}
\end{figure}

In figure~\ref{fig:cov-noncov} we compare the covariant
[eqs~\eqref{l1-cov}, \eqref{l2-cov}] to the non-covariant
[eqs~\eqref{l1-noncov}, \eqref{l2-noncov}] extraction of $\lambda_1$
and $\lambda_2$, for the H06 and L07 ensembles, with and without
tree-level correction.  We see that the results for $\lambda_1$ agree perfectly in the
infrared, but diverge for $pa\gtrsim0.5$.  This is in part due to the
tree-level correction: the uncorrected (raw) data agree for
$pa\lesssim1$, but the non-covariant $\lambda_1$ is brought down by
the tree-level correction, while the covariant $\lambda_1$ is brought
up.  In the ultraviolet, the covariant $\lambda_1$ goes to zero at tree level, and
this is also seen in the raw data.  The tree-level corrected,
covariant $\lambda_1$ therefore becomes very noisy in the UV.

In the case of $\lambda_2$, the covariant and non-covariant
determinations agree very well after tree-level corrections for
$pa\lesssim1$, while in the ultraviolet the covariant data fall off
somewhat more slowly than the non-covariant ones.

In figure~\ref{fig:mass-flavour} we compare the results for the form
factors $\lambda_1,\lambda_2,\lambda_3$, renormalised at 2\,GeV, for
the Q07 (quenched), H07
and L07 ensembles.  All these have the same lattice spacings, but
differ in their quark flavour content.  In the left panel, comparing
the Q07 to the H07 ensemble, we see that the dynamical fermions have a
moderate effect on all three form factors, and that including the
sea quarks tends to enhance all of the form factors in the infrared.
In the right panel we compare the H07 and L07 ensembles, which differ
only by their quark mass.  We see a similar effect, where the lighter
quark gives a stronger enhancement in the infrared.

\begin{figure}[tbp]
\includegraphics[width=0.48\textwidth]{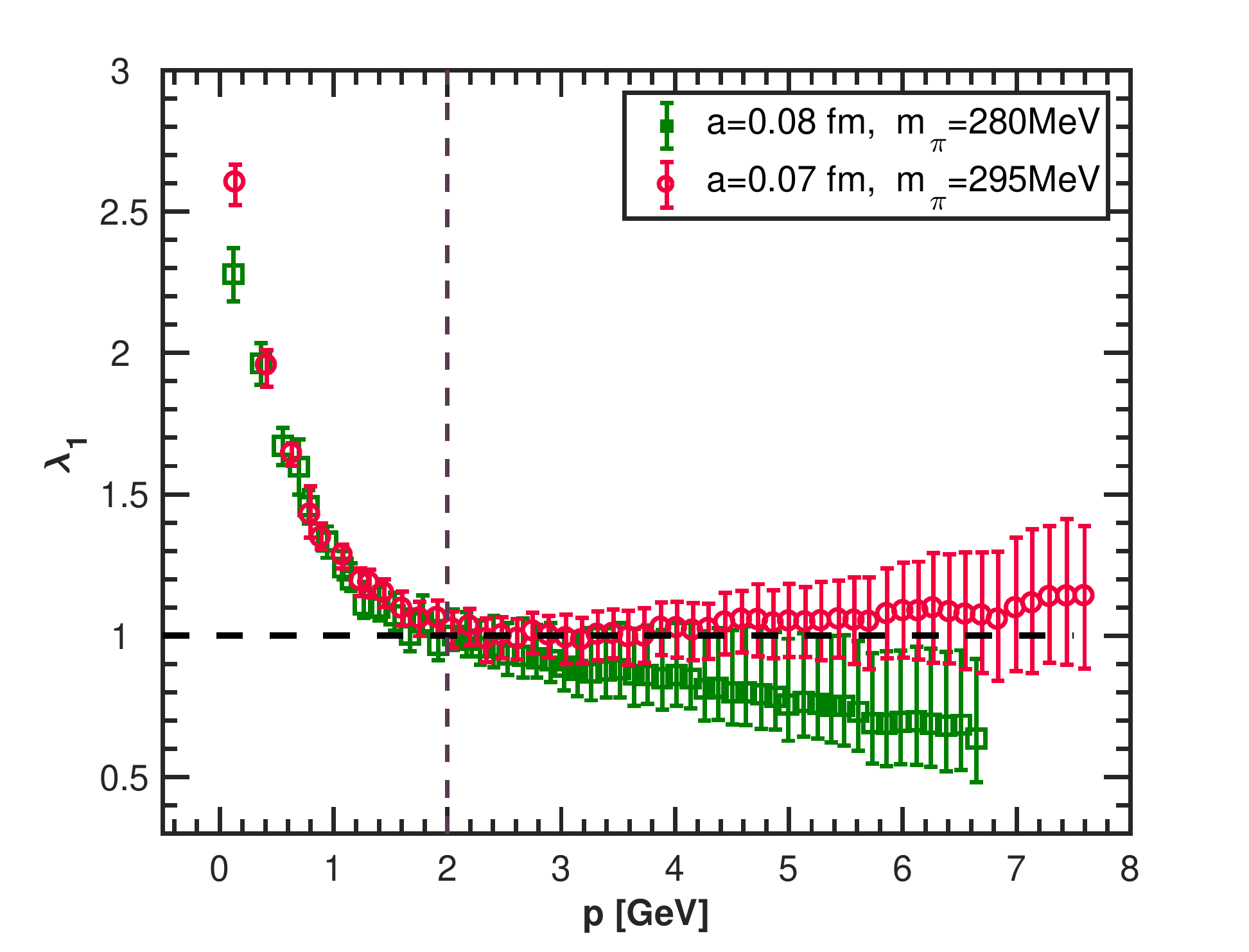}
\includegraphics[width=0.48\textwidth]{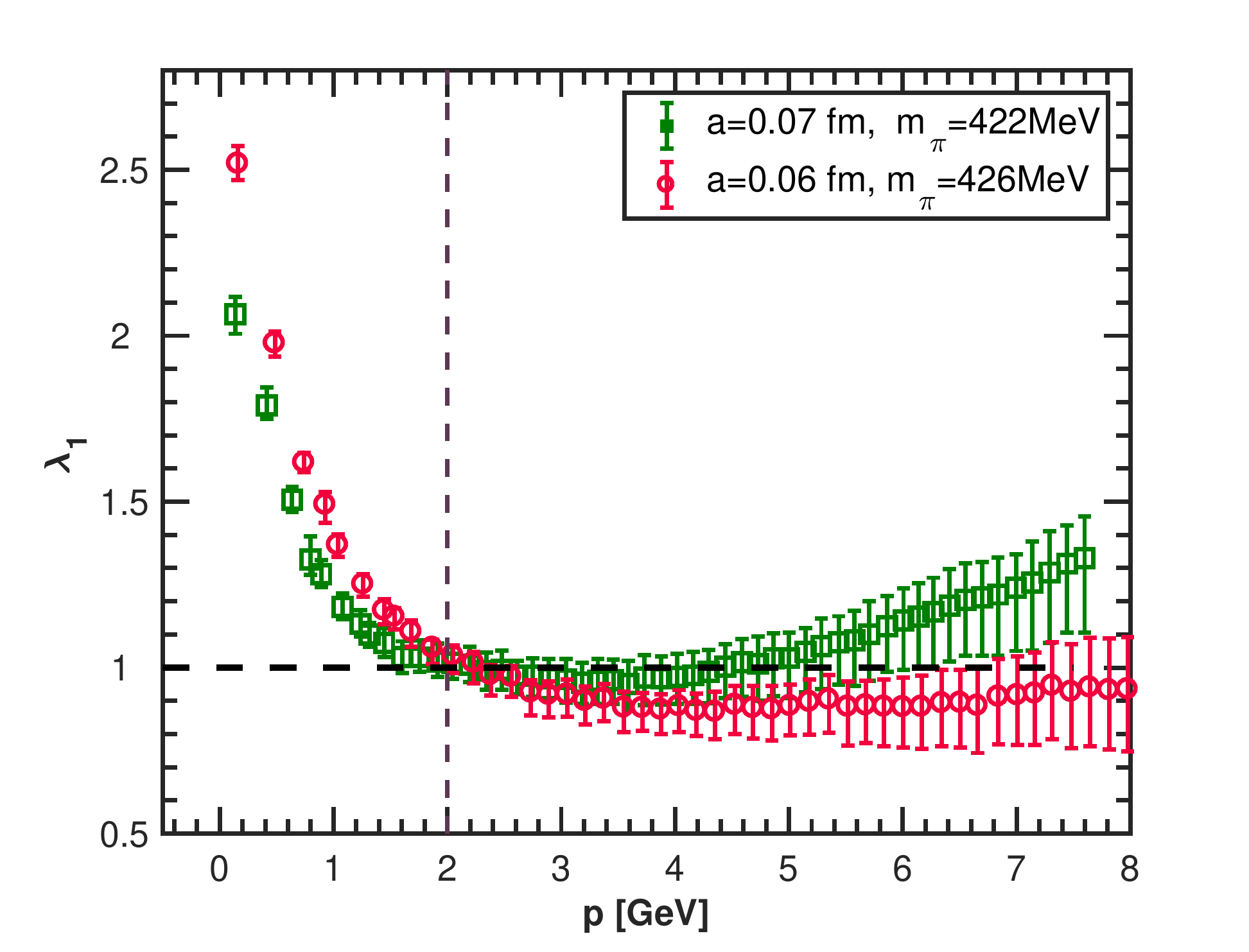}\\
\includegraphics[width=0.48\textwidth]{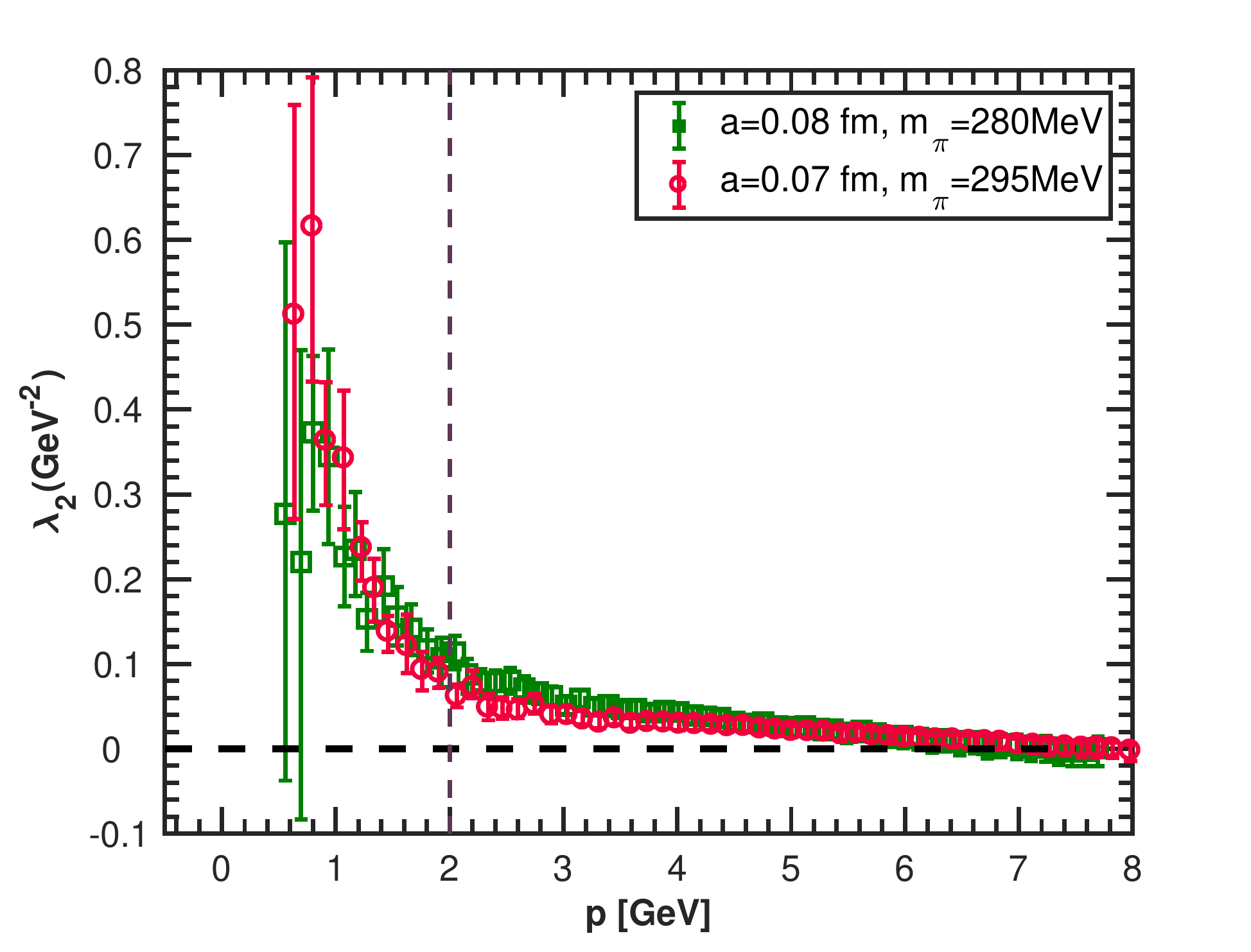}
\includegraphics[width=0.48\textwidth]{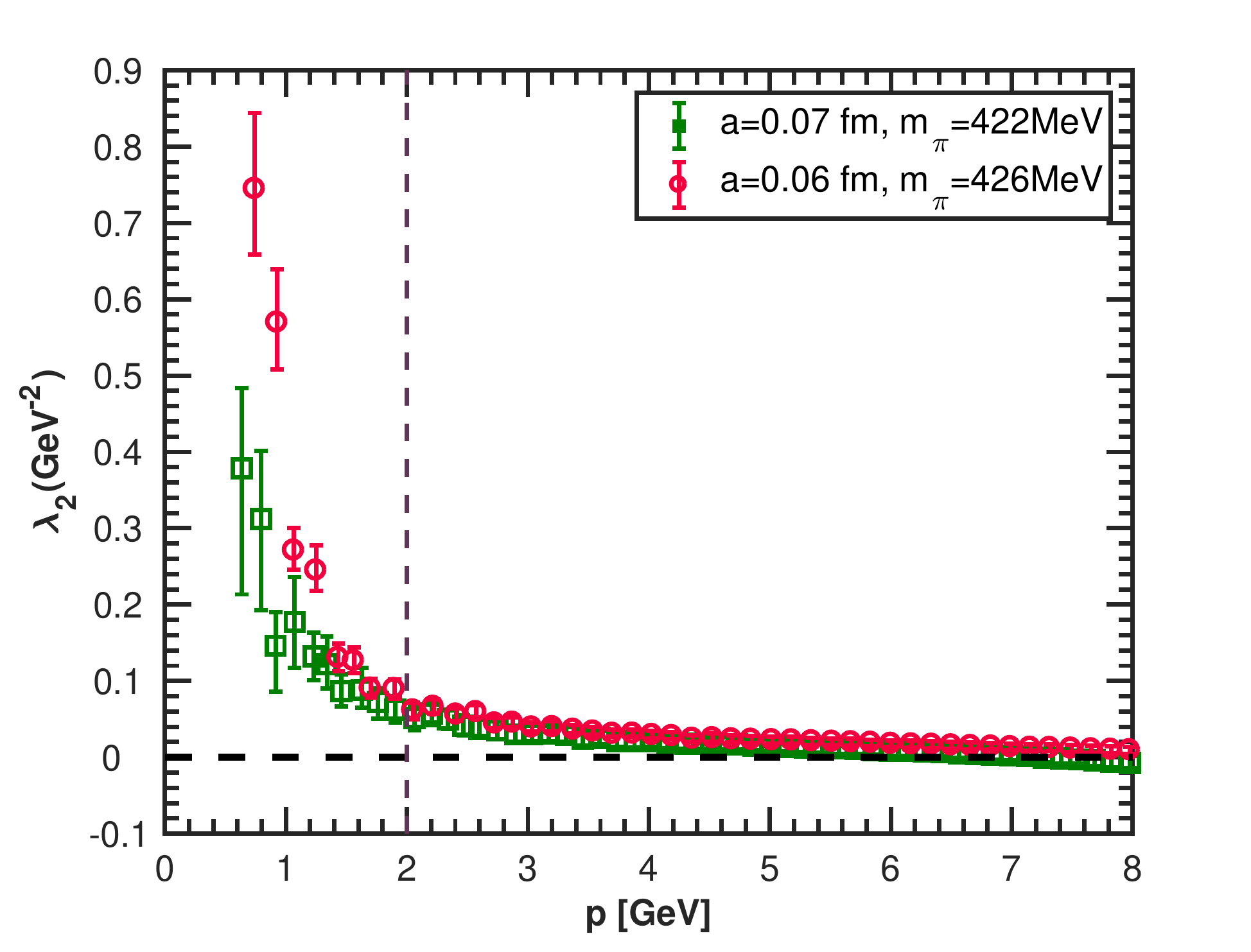}\\
\includegraphics[width=0.48\textwidth]{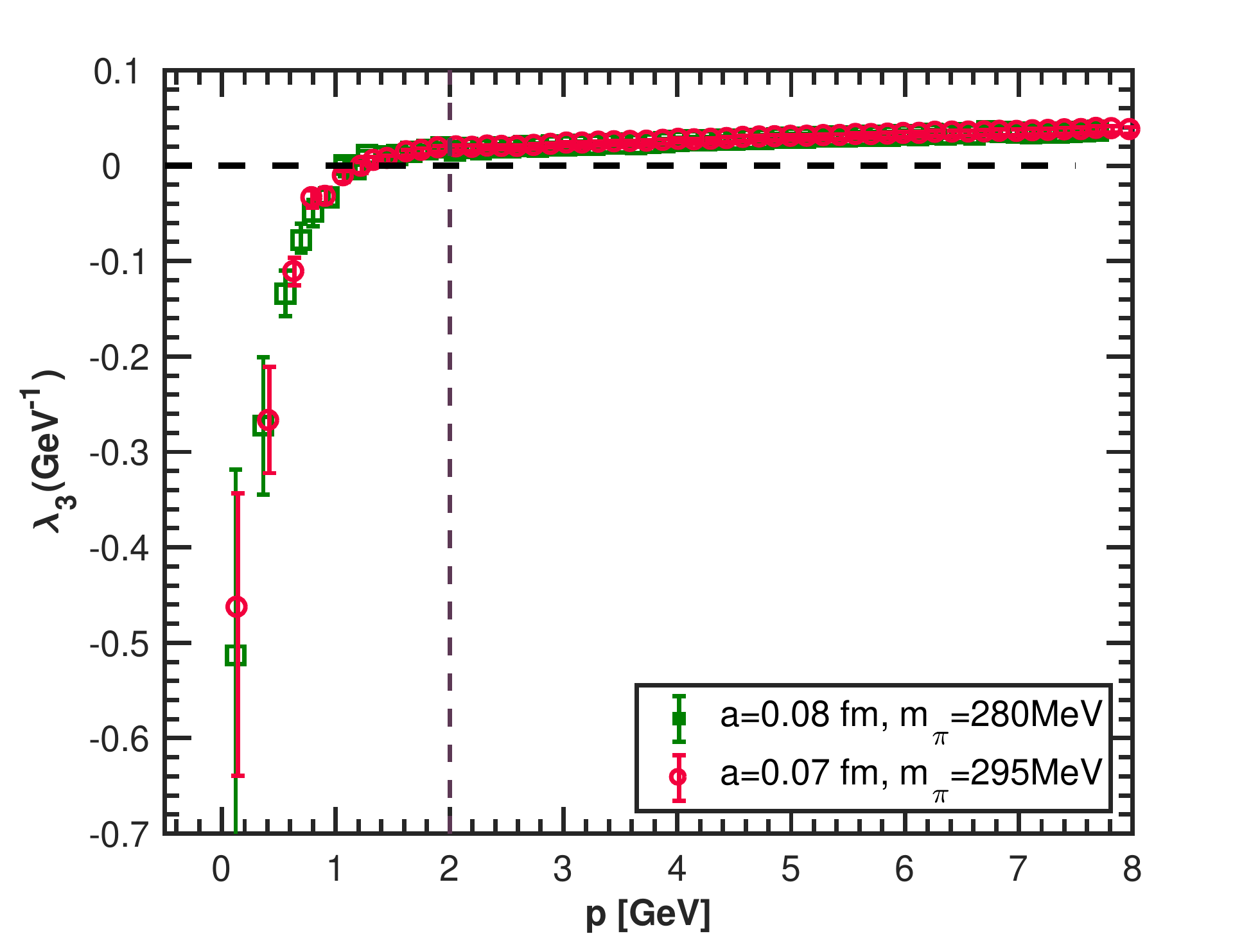}
\includegraphics[width=0.48\textwidth]{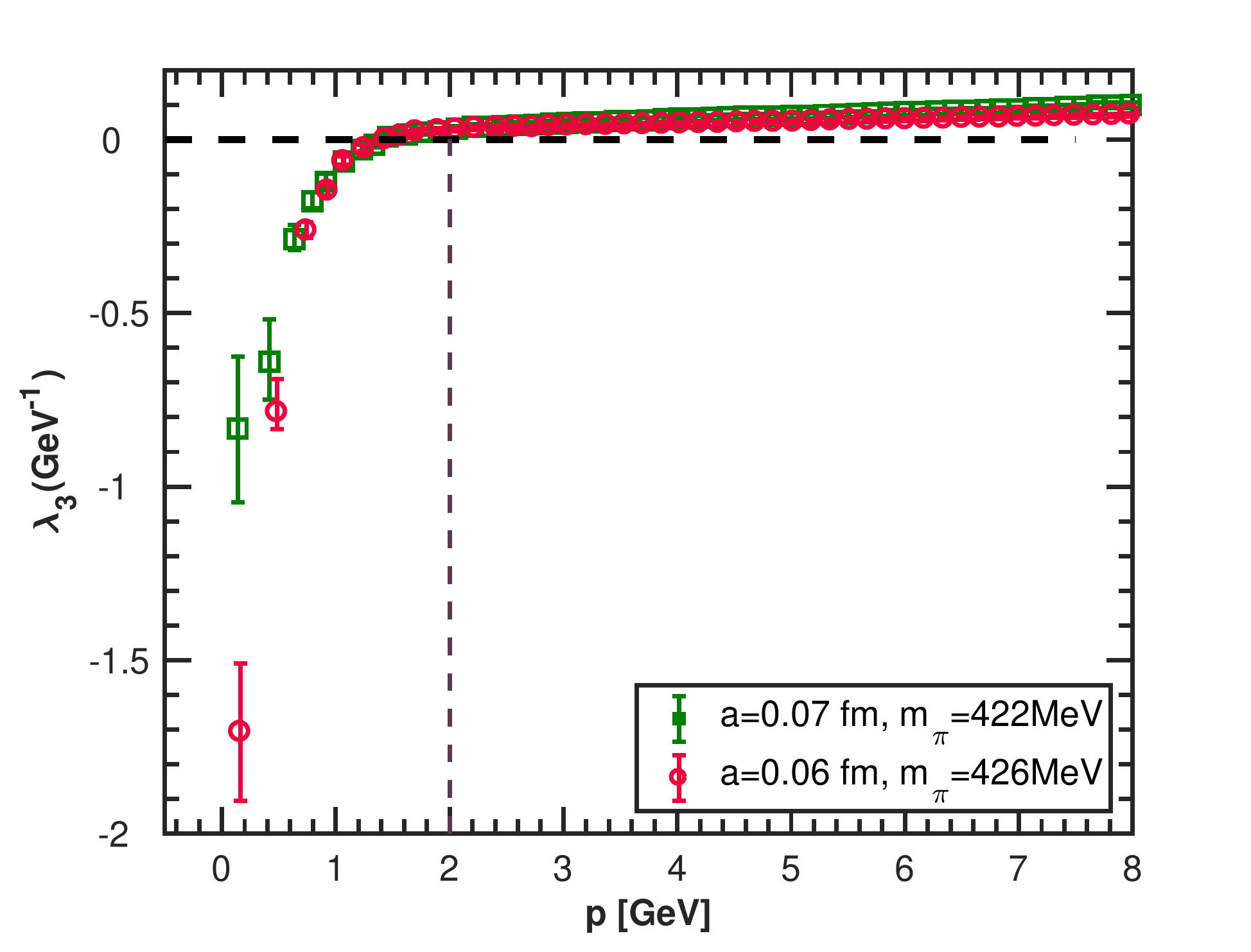}
\caption{Lattice spacing dependence of (top to bottom) $\lambda_1,
  \lambda_2$ and $\lambda_3$.  The left panel
  compares the L08 and L07 ensembles, while the
  right panel compares the H07 and H06 ensembles.}
\label{fig:latspacing}
\end{figure}
Next, in figure~\ref{fig:latspacing} we compare the form factors at
different lattice spacings for the same physical quark mass.  For the
lighter quarks (L07 and L08) we see very little lattice spacing
dependence, while for the heavier quarks (H06 and H07) a clear
enhancement in the infrared is seen for all form factors as we
approach the continuum limit.

\begin{figure}[tbh]
\includegraphics[width=0.48\textwidth]{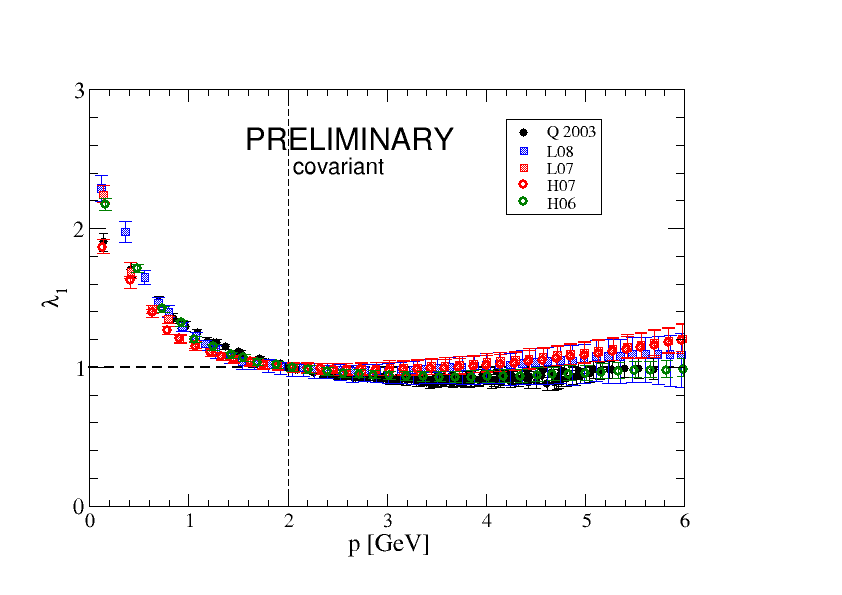}\includegraphics[width=0.48\textwidth]{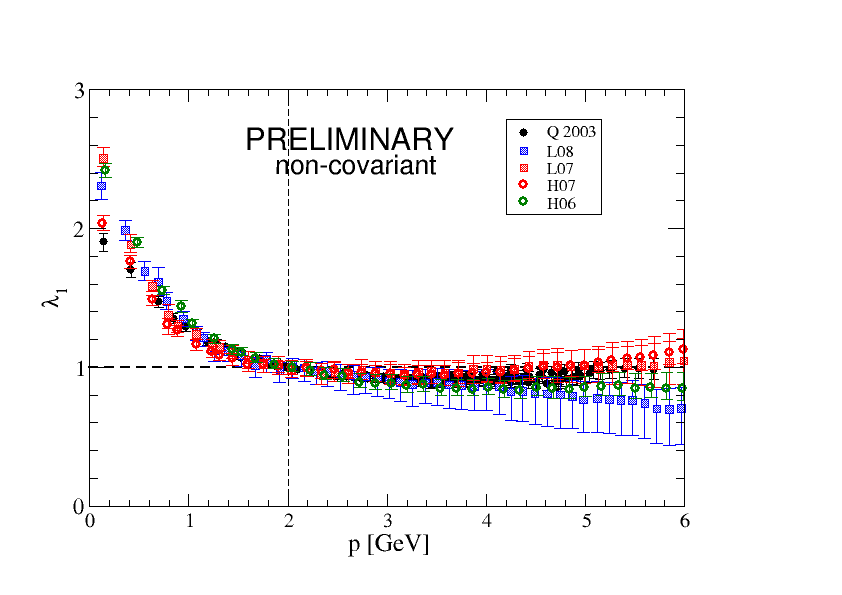}\\
\includegraphics[width=0.48\textwidth]{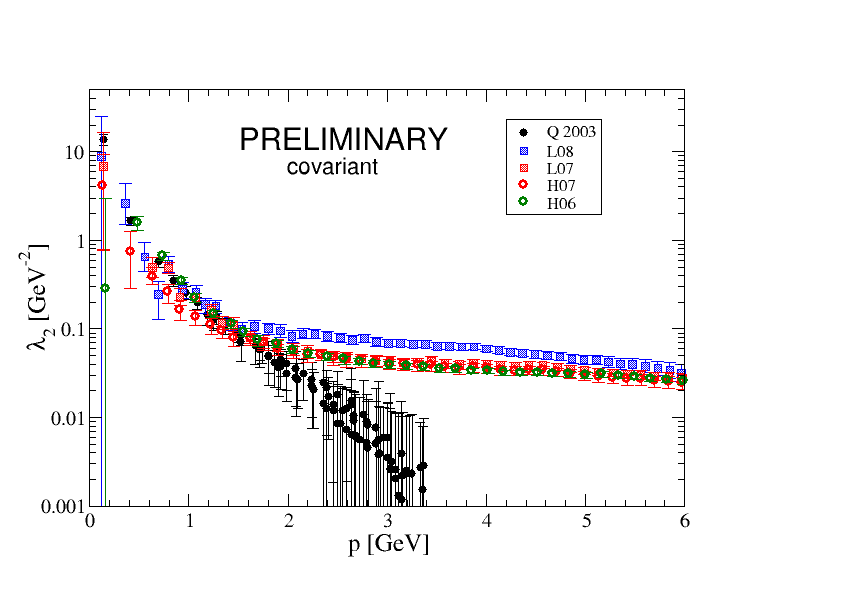}\includegraphics[width=0.48\textwidth]{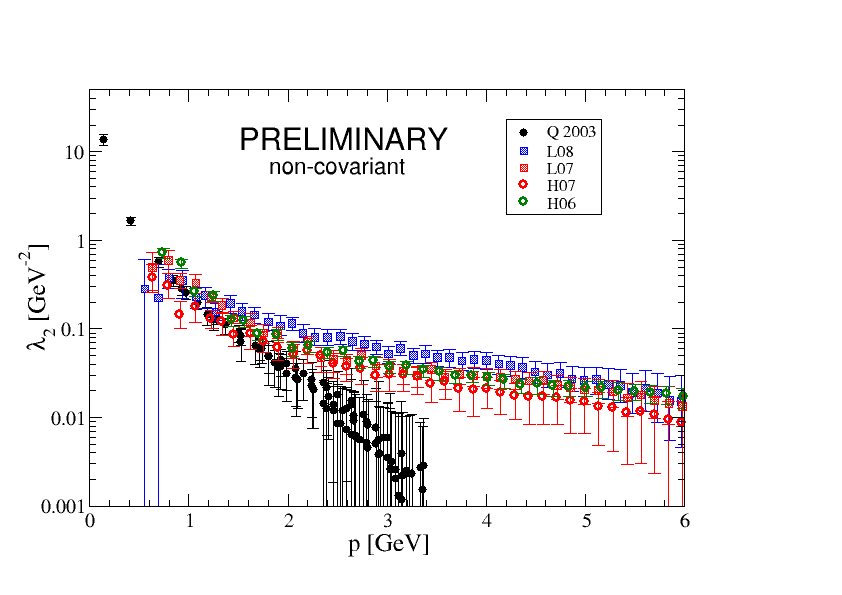}\\
\includegraphics[width=0.48\textwidth]{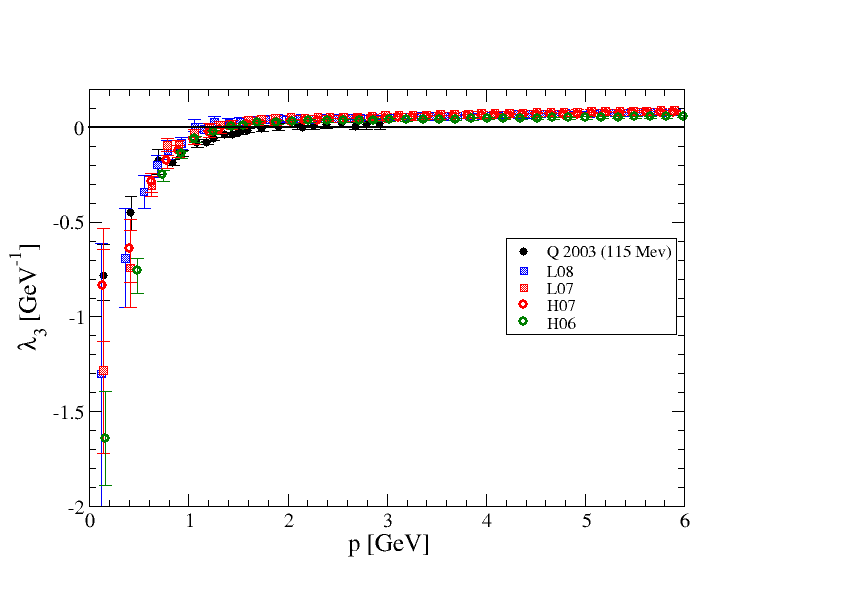}
\caption{All form factors, renormalised at $\mu=2\,$GeV, compared with
  the quenched results from 2003 \cite{Skullerud:2003qu}, labelled ``Q
  2003''.  Left: covariant extraction; right: non-covariant extraction.}
\label{fig:compare}
\end{figure}
Finally, in figure~\ref{fig:compare} we show the results from all our
$N_f=2$ ensembles, together with the earlier quenched results from
\cite{Skullerud:2003qu}.  The earlier results were obtained using a
mean-field (tadpole) improved rather than a nonperturbatively determined
clover coefficient, with a lattice spacing of 0.093\,fm and a valence
quark mass corresponding to $m_\pi\approx880\,$MeV
\cite{UKQCD:1999ukw}.  We find good agreement with the old results,
suggesting once again that quenching, lattice spacing and quark mass
effects are moderate.

\section{Outlook}

We have presented results for the quark--gluon vertex in the soft
gluon limit with two flavours of $\order(a)$-improved Wilson
fermion.  The results are consistent with previous results in the
quenched approximation, showing moderate dependence on flavour
content, quark mass and lattice spacing in the range studied here.  We
have employed two different methods (covariant and non-covariant) to
extract the form factors $\lambda_1$ and $\lambda_2$, and the results
are consistent in the infrared.  There is some discrepancy between the
two methods for $\lambda_1$ at mid-momentum which still needs to be
understood.

Our next step will be to compute the vertex in more general
kinematics, which will give access to the full set of 12 form factors.
Recent studies based on Dyson--Schwinger equations and exploiting
Slavnov--Taylor identities \cite{Binosi:2016wcx,Gao:2021wun} suggest
that only a few of these form factors give a significant contribution
to the strength of the vertex, and our calculations will aim to check
these findings.

\section*{Acknowledgments}

The gauge fixing and calculations of the fermion propagators were
performed on the HLRN supercomputing facilities in Berlin and Hanover.
JIS has been supported by Science Foundation Ireland grant
11/RFP.1/PHY/1362. AS acknowledges support by the DFG as member of the SFB/TRR55 and GRK1523. OO and PJS acknowledge support from
FCT (Portugal) Projects No. UID/FIS/04564/2019 and UID/FIS/04564/2020.
PJS acknowledges financial support from FCT (Portugal) under Contract
No. CEECIND/00488/2017.
AK thanks Prof. A.W. Thomas for supporting this work.

\bibliographystyle{JHEP-2}
\bibliography{qgv_references}

\providecommand{\href}[2]{#2}\begingroup\raggedright\begin{thebibliography}{10}

\bibitem{Williams:2015cvx}
R.~Williams, C.~S. Fischer and W.~Heupel, {\it Light mesons in {QCD} and
  unquenching effects from the {3PI} effective action},  {\em Phys. Rev. D}
  {\bf 93} (2016), no.~3 034026 [\href{http://arXiv.org/abs/1512.00455}{{\tt
  arXiv:1512.00455}}].

\bibitem{Binosi:2016wcx}
D.~Binosi, L.~Chang, J.~Papavassiliou, S.-X. Qin and C.~D. Roberts, {\it
  Natural constraints on the gluon-quark vertex},  {\em Phys. Rev.} {\bf D95}
  (2017), no.~3 031501(R) [\href{http://arXiv.org/abs/1609.02568}{{\tt
  arXiv:1609.02568}}].

\bibitem{Aguilar:2018epe}
A.~C. Aguilar, J.~C. Cardona, M.~N. Ferreira and J.~Papavassiliou, {\it Quark
  gap equation with non-abelian {B}all--{C}hiu vertex},  {\em Phys. Rev.} {\bf
  D98} (2018), no.~1 014002 [\href{http://arXiv.org/abs/1804.04229}{{\tt
  arXiv:1804.04229}}].

\bibitem{Gao:2021wun}
F.~Gao, J.~Papavassiliou and J.~M. Pawlowski, {\it Fully coupled functional
  equations for the quark sector of {QCD}},  {\em Phys. Rev. D} {\bf 103}
  (2021), no.~9 094013 [\href{http://arXiv.org/abs/2102.13053}{{\tt
  arXiv:2102.13053}}].

\bibitem{Albino:2021rvj}
L.~Albino, A.~Bashir, B.~El-Bennich, E.~Rojas, F.~E. Serna and R.~C.
  da~Silveira, {\it The impact of transverse {S}lavnov--{T}aylor identities on
  dynamical chiral symmetry breaking},
  \href{http://arXiv.org/abs/2108.06204}{{\tt arXiv:2108.06204}}.

\bibitem{Skullerud:2002ge}
J.-I. Skullerud and A.~K{\i}z{\i}lers{\"u}, {\it Quark-gluon vertex from
  lattice {QCD}},  {\em JHEP} {\bf 09} (2002) 013
  [\href{http://arXiv.org/abs/hep-ph/0205318}{{\tt hep-ph/0205318}}].

\bibitem{Skullerud:2003qu}
J.-I. Skullerud, P.~O. Bowman, A.~K{\i}z{\i}lers{\"u}, D.~B. Leinweber and
  A.~G. Williams, {\it Nonperturbative structure of the quark gluon vertex},
  {\em JHEP} {\bf 04} (2003) 047
  [\href{http://arXiv.org/abs/hep-ph/0303176}{{\tt hep-ph/0303176}}].

\bibitem{Lin:2005zd}
H.-W. Lin, {\it Quark-gluon vertex with an off-shell {$O(a)$}-improved chiral
  fermion action},  {\em Phys. Rev. D} {\bf 73} (2006) 094511
  [\href{http://arXiv.org/abs/hep-lat/0510110}{{\tt arXiv:hep-lat/0510110}}].

\bibitem{Kizilersu:2006et}
A.~K{\i}z{\i}lers{\"u}, D.~B. Leinweber, J.-I. Skullerud and A.~G. Williams,
  {\it Quark-gluon vertex in general kinematics},  {\em Eur. Phys. J. C} {\bf
  50} (2007) 871--875 [\href{http://arXiv.org/abs/hep-lat/0610078}{{\tt
  arXiv:hep-lat/0610078}}].

\bibitem{Kizilersu:2021jen}
A.~K\i{}z\i{}lers\"u, O.~Oliveira, P.~J. Silva, J.-I. Skullerud and
  A.~Sternbeck, {\it Quark-gluon vertex from {$N_f=2$} lattice {QCD}},  {\em
  Phys. Rev. D} {\bf 103} (2021), no.~11 114515
  [\href{http://arXiv.org/abs/2103.02945}{{\tt arXiv:2103.02945}}].

\bibitem{Bali:2014gha}
G.~S. Bali, S.~Collins, B.~Gl{\"a}{\ss}le, M.~G{\"o}ckeler, J.~Najjar, R.~H.
  R{\"o}dl, A.~Sch{\"a}fer, R.~W. Schiel, A.~Sternbeck and W.~S{\"o}ldner, {\it
  The moment {$\langle x\rangle_{u-d}$} of the nucleon from {$N_f=2$} lattice
  {QCD} down to nearly physical quark masses},  {\em Phys. Rev.} {\bf D90}
  (2014), no.~7 074510 [\href{http://arXiv.org/abs/1408.6850}{{\tt
  arXiv:1408.6850}}].

\bibitem{UKQCD:1999ukw}
{\bf UKQCD} Collaboration, K.~C. Bowler {\em et.~al.}, {\it Quenched {QCD} with
  {$O(a)$} improvement. 1. the spectrum of light hadrons},  {\em Phys. Rev. D}
  {\bf 62} (2000) 054506 [\href{http://arXiv.org/abs/hep-lat/9910022}{{\tt
  arXiv:hep-lat/9910022}}].

\end{thebibliography}\endgroup
\end{document}